\documentclass[twocolumn,pra,aps]{revtex4}
\usepackage{amsmath}
\usepackage{amssymb}
\usepackage{epsfig}
\usepackage{bm}
\usepackage{epsfig}

\begin{document}

\title{Condensation of N interacting bosons: Hybrid approach to condensate
fluctuations}
\author{Anatoly A. Svidzinsky and Marlan O. Scully}
\affiliation{{\small Institute for Quantum Studies and Department of Physics, Texas A\&M
Univ., Texas 77843 }\\
Applied Physics and Materials Science Group, Eng. Quad., Princeton Univ.,
Princeton 08544 }
\date{\today }

\begin{abstract}
We present a new method of calculating the distribution function and
fluctuations for a Bose-Einstein condensate (BEC) of N interacting atoms.
The present formulation combines our previous master equation and canonical
ensemble quasiparticle techniques. It is applicable both for ideal and
interacting Bogoliubov BEC and yields remarkable accuracy at all
temperatures. For the interacting gas of 200 bosons in a box we plot the 
temperature
dependence of the first four central moments of the condensate particle
number and compare the results with the ideal gas. For the interacting
mesoscopic BEC, as with the ideal gas, we find a smooth transition for the
condensate particle number as we pass through the critical temperature.
\end{abstract}

\maketitle

Bose-Einstein condensation (BEC) near the critical temperature $T_{c}$ is a
fascinating subject. Indeed the time line from the early studies of
Uhlenbeck \cite{U27} concerning the unphysical cusp at $T_{c}$ in Einstein's
analysis, to the \textquotedblleft Grand canonical catastrophe" \cite{HKK},
arising from ill fated attempts to describe fluctuations \cite{Koch06},
covers most of the twentieth century. Naturally, a major stimulus for the
present work is the pioneering BEC experiments in liquid He-II \cite{rep}
and in laser cooled gases \cite{Meys01}. The attendant theoretical advances
in the study of BEC in ideal and interacting gases likewise constitute a
rich field \cite{Kett96,la}.

Furthermore, subtle issues such as the statistics of condensate atoms in a
mesoscopic system of $N\sim 10^{2}-10^{3}$ particles are now of interest.
Condensate fluctuations can be measured by means of a scattering of series
of short laser pulses \cite{Idzi00}, see also \cite{Chuu05}. We note that
the BEC is often referred to as an atom laser; indeed the problem of BEC
statistics near $T_{c}$ is analogous to studying the photon statistics of
the laser in the passage from below to above threshold \cite{KSZZ,Imam97}.

In the present Letter, we give, for the first time, a simple, surprisingly
accurate account of fluctuations in an interacting Bogoliubov gas valid for
all temperatures. We omit, however, effects of interaction between
Bogoliubov quasiparticles since we treat a weakly interacting gas. The
analysis is based on a master equation approach deriving from a union of the
ideal gas (in the spirit of the quantum theory of the laser) BEC treatment
of the first two papers in this series on the condensation of N bosons \cite%
{KSZZ} and the results for the interacting gas BEC obtained based on the
canonical ensemble quasiparticle formalism \cite{KKS-PRL}.

We emphasize that although useful papers have been published dealing with
various limiting cases, so far there has been no treatment of this problem
valid at all temperatures \cite{Koch06}. Ref. \cite{KKS-PRL} presented
analytical formulas for all moments of the condensate particle number
fluctuations in the weakly interacting Bose gas. However, \cite{KKS-PRL} and
other approaches (e.g. Ref. \cite{Pit98}) are only valid provided the
average number of condensate particles is much larger than its variance.
However, near and above $T_{c}$ this is not true, and this causes the
failure of such treatments.

\begin{figure}[h]
\bigskip 
\centerline{\epsfxsize=0.52\textwidth\epsfysize=0.4\textwidth
\epsfbox{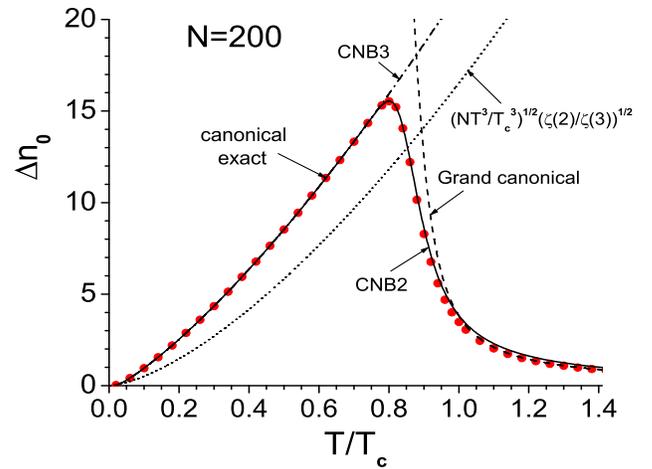}}
\caption{The variance for the condensate particle number as a function of
temperature for an ideal gas of $N=200$ particles in an isotropic harmonic
trap obtained by the master equation approach of \protect\cite{KSZZ} (solid
line), ``exact" numerical simulations in the canonical ensemble (dots) and
the grand canonical answer (dashed line). Result of Refs. \protect\cite%
{HKK,KKS-PRL} is plotted as dash-dot line. Small dots show the thermodynamic
limit formula of Politzer \protect\cite{Politzer}.}
\label{v2}
\end{figure}

For an ideal Bose gas the master equation approach of \cite{KSZZ} naturally
includes the N particle constraint and provides an analytical solution for
the partition function and fluctuations accurate at all temperatures, as
shown in Fig. \ref{v2}. For an interacting gas, the problem of fluctuations
is rather delicate. In a recent paper \cite{Scul06} we gave a preliminary
master equation analysis for N interacting atoms which accurately described
the average number of particles in the condensate. However the fluctuations
were handled less well. The present analysis gives all central moments with
remarkable accuracy when compared to \cite{KKS-PRL} below $T_{c}$. Above $%
T_{c}$ the present results go smoothly into the ideal gas limit as they must.

%%%%%%%%%%%%%%%%%%%% Fig 1 %%%%%%%%%%%%%%%%%%%%%%%%%%%%%
%%%%%%%%%%%%%%%%%%%%%%%%%%%%%%%%%%%%%%%%%%%%%%%%%%%%%%%%%

The central tool used in the ideal gas analysis of \cite{KSZZ} and the
interacting gas study of \cite{Scul06} was the laser-like master equation
for the probability $p_{n_{0}}$ of finding $n_{0}$ atoms in the condensate,
given that there are $N$ total particles%
\begin{equation*}
\frac{1}{\kappa }\dot{p}%
_{n_{0}}=-K_{n_{0}}(n_{0}+1)p_{n_{0}}+K_{n_{0}-1}n_{0}p_{n_{0}-1}-
\end{equation*}%
\begin{equation}
H_{n_{0}}n_{0}p_{n_{0}}+H_{n_{0}+1}(n_{0}+1)p_{n_{0}+1},
\end{equation}%
where $\kappa $ is an uninteresting rate constant, $H_{n_{0}}$ and $%
K_{n_{0}} $ are heating and cooling coefficients. In equilibrium the rates
of any two opposite processes are equal to each other, e.g., $%
K_{n_{0}}(n_{0}+1)p_{n_{0}}=H_{n_{0}+1}(n_{0}+1)p_{n_{0}+1}$. The detailed
balance condition yields%
\begin{equation}
\frac{p_{n_{0}+1}}{p_{n_{0}}}=\frac{K_{n_{0}}}{H_{n_{0}+1}}.  \label{p1}
\end{equation}

Since the occupation number of the ground state cannot be larger than $N$
there is a canonical ensemble constraint $p_{N+1}=0$ and, hence, $K_{N}=0$.
In contrast to $p_{n_{0}}$, the ratio $p_{n_{0}+1}/p_{n_{0}}$ as a function
of $n_{0}$ shows simple monotonic behavior. We approximate $K_{n_{0}}$ and $%
H_{n_{0}}$ by a few terms of the Taylor expansion near the point $n_{0}=N$%
\begin{equation}
K_{n_{0}}=(N-n_{0})(1+\eta )+\alpha (N-n_{0})^{2},  \label{p2}
\end{equation}%
\begin{equation}
H_{n_{0}}=\mathcal{H}+(N-n_{0})\eta +\alpha (N-n_{0})^{2}.  \label{p3}
\end{equation}%
Parameters $\mathcal{H}$, $\eta $ and $\alpha $ are independent of $n_{0}$;
they are functions of the occupation of the excited levels. We derive them
below by matching the first three central moments in the low temperature
limit with the result of \cite{KKS-PRL}. We note that the detailed balance
equation (\ref{p1}) is the Pad\'{e} approximation \cite{Pade} of the
function $p_{n_{0}+1}/p_{n_{0}}$. Pad\'{e} summation has proven to be useful
in many applications, including condensed-matter problems and quantum field
theory.

Eqs. (\ref{p1})-(\ref{p3}) yield an analytical expression for the condensate
distribution function 
\begin{equation}
p_{n_{0}}=\frac{1}{{\mathcal{Z}_{N}}}\frac{%
(N-n_{0}-1+x_{1})!(N-n_{0}-1+x_{2})!}{(N-n_{0})!(N-n_{0}+(1+\eta )/\alpha )!}%
,  \label{p4}
\end{equation}%
where $x_{1,2}=(\eta \pm \sqrt{\eta ^{2}-4\alpha \mathcal{H}})/2\alpha $ and 
$\mathcal{Z}_{N}$ is the normalization constant determined by $%
\sum\limits_{n_{0}=0}^{N}$ $p_{n_{0}}=1$. In the particular case $\eta
=\alpha =0$ Eq. (\ref{p4}) reduces to 
\begin{equation}
p_{n_{0}}=\frac{1}{Z_{N}}\frac{\mathcal{H}^{N-n_{0}}}{(N-n_{0})!},
\label{p5}
\end{equation}%
where $Z_{N}=e^{\mathcal{H}}\Gamma (N+1,\mathcal{H})/N!$ is the partition
function and $\Gamma $ is an incomplete gamma-function. For an ideal gas Eq.
(\ref{p5}) describes accurately the condensate statistics at low temperature 
\cite{KSZZ}. The statistics is not Poissonian $p_{n}=\bar{n}^{n}e^{-\bar{n}%
}/n!$, as would be expected for a coherent state.

Using the distribution function (\ref{p4}) we find that, in the validity
range of \cite{KKS-PRL} (at low $T$), the first three central moments $\mu
_{m}\equiv <(n_{0}-\bar{n}_{0})^{m}>$ are 
\begin{equation}
\bar{n}_{0}=N-\mathcal{H},\qquad \mu _{2}=(1+\eta )\mathcal{H}+\alpha 
\mathcal{H}^{2},  \label{p6}
\end{equation}%
\begin{equation}
\mu _{3}=-\mathcal{H}(1+\eta +\alpha \mathcal{H})(1+2\eta +4\alpha \mathcal{H%
}).  \label{p8}
\end{equation}%
Eqs. (\ref{p6}), (\ref{p8}) thus yield%
\begin{equation}
\mathcal{H}=N-\bar{n}_{0},\qquad \eta =\frac{1}{2}\left( \frac{\mu _{3}}{\mu
_{2}}-3+\frac{4\mu _{2}}{\mathcal{H}}\right) ,  \label{p9}
\end{equation}%
\begin{equation}
\alpha =\frac{1}{\mathcal{H}}\left( \frac{1}{2}-\frac{\mu _{2}}{\mathcal{H}}-%
\frac{\mu _{3}}{2\mu _{2}}\right) .  \label{p11}
\end{equation}

On the other hand, the result of \cite{KKS-PRL} for an interacting
Bogoliubov gas is (see also \cite{Pit98} for $\bar{n}_{0}$ and $\mu _{2}$) 
\begin{equation}
\bar{n}_{0}=N-\sum\limits_{k\neq 0}\left[ \left( u_{k}^{2}+v_{k}^{2}\right)
f_{k}+v_{k}^{2}\right] ,  \label{a1}
\end{equation}%
\begin{equation}
\mu _{2}=\sum\limits_{k\neq 0}\left[
(1+8u_{k}^{2}v_{k}^{2})(f_{k}^{2}+f_{k})+2u_{k}^{2}v_{k}^{2}\right] ,
\label{a2}
\end{equation}%
\begin{equation*}
\mu _{3}=-\sum\limits_{k\neq 0}\left( u_{k}^{2}+v_{k}^{2}\right) \left[
(1+16u_{k}^{2}v_{k}^{2})(2f_{k}^{3}+3f_{k}^{2}+f_{k})+\right.
\end{equation*}%
\begin{equation}
\left. 4u_{k}^{2}v_{k}^{2}(1+2f_{k})\right] ,  \label{a3}
\end{equation}%
where $f_{k}=1/[\exp (E_{k}/k_{B}T)-1]$ is the number of elementary
excitations with energy $E_{k}$ present in the system at thermal
equilibrium, $u_{k}$ and $v_{k}$ are Bogoliubov amplitudes. Substitute for $%
\bar{n}_{0}$, $\mu _{2}$ and $\mu _{3}$ in Eqs. (\ref{p9}), (\ref{p11})
their expressions of Ref. \cite{KKS-PRL} (\ref{a1})-(\ref{a3}) yields the
unknown parameters $\mathcal{H}$, $\eta $ and $\alpha $. The beauty of our
\textquotedblleft matched asymptote" derivation is that the formulas for $%
\mathcal{H}$, $\eta $ and $\alpha $\ are applicable at all temperatures,
i.e. not only in the validity range of \cite{KKS-PRL}. The distribution
function (\ref{p4}) together with Eqs. (\ref{p9}), (\ref{p11}) provides
complete knowledge of the condensate statistics at all $T$. Taking $v_{k}=0$
and $u_{k}=1$ in (\ref{a1})-(\ref{a3}) we obtain the ideal gas limit.

\begin{figure}[h]
\bigskip 
\centerline{\epsfxsize=0.52\textwidth\epsfysize=0.4\textwidth
\epsfbox{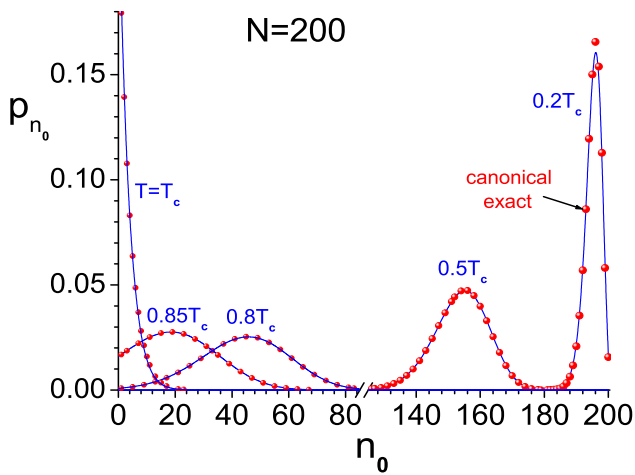}}
\caption{Distribution function for the condensate particle number $n_0$ at
different temperature obtained by the present approach (solid lines) and
``exact" numerical simulations in the canonical ensemble (dots). The results
are obtained for an ideal gas of $N=200$ particles in an isotropic harmonic
trap.}
\label{pn0}
\end{figure}

Next we test our method for an ideal gas (in a harmonic trap). In this case
an \textquotedblleft exact" numerical simulation in the canonical ensemble 
\cite{la} is available for comparison. Results of such simulation are shown
by dots in Figs. \ref{pn0} and \ref{n14trap}. In Fig. \ref{pn0} we plot the
distribution function for the number of atoms in condensate at different
temperatures and $N=200$. At $T\ll T_{c}$ the distribution shows a sharp
peak near $\bar{n}_{0}$ and becomes broader at higher $T$. The present Eq. (%
\ref{p4}) (solid line) yields excellent agreement with the \textquotedblleft
exact" dots at all temperatures. Fig. \ref{n14trap} shows the average
condensate particle number $\bar{n}_{0}$, its variance, third and fourth
central moments $\mu _{m}$ and fourth cumulant $\kappa _{4}$ \cite{kum} as a
function of $T$ for $N=200$ particles in a harmonic trap. Solid lines are
the result of the present approach (we call it CNB5 \cite{note}) which is in
remarkable agreement with the \textquotedblleft exact" dots at all
temperatures both for $\mu _{m}$ and $\kappa _{4}$. Central moments and
cumulants higher than fourth order are not shown here, but they are also
remarkably accurate at all temperatures. Results of \cite{KKS-PRL} are
given by dashed lines which are accurate only at sufficiently low $T$.
Deviation of higher order cumulants ($m=3,4,\ldots $) from zero indicates
that the fluctuations are not Gaussian.

%%%%%%%%%%%%%%%%%%%% Fig 2 %%%%%%%%%%%%%%%%%%%%%%%%%%%%%
%%%%%%%%%%%%%%%%%%%%%%%%%%%%%%%%%%%%%%%%%%%%%%%%%%%%%%%%%

Clearly our method passes the ideal gas test with flying colors. Please note
the excellent agreement with the exact analysis for the third central moment
and fourth cumulant $\kappa _{4}$ given in Fig. \ref{n14trap}.

Next we apply the present technique to $N$ interacting Bogoliubov particles
confined in a box of volume $V$. The interactions are characterized by the
gas parameter $an^{1/3} $, where $a$ is the s-wave scattering length and $%
n=N/V$ is the particle density. The energy of Bogoliubov quasiparticles $%
E_{k}$ depends on $\bar{n}_{0}$, hence, the equation $\bar{n}%
_{0}=\sum\limits_{n_{0}=0}^{N}$ $n_{0}p_{n_{0}}$ for $\bar{n}_{0}$ must be
solved self-consistently. In Fig. \ref{n14box} we plot $\bar{n}_{0}$, the
variance $\Delta n_{0}$, third and fourth central moments as a function of $%
T $ for an ideal and interacting ($an^{1/3}=0.1$) gas in the box. Solid
lines show the result of the present approach, while \cite{KKS-PRL} is
represented by dashed lines. The present results agree well for all $\mu
_{m} $ with \cite{KKS-PRL} in the range of its validity. Near and above $%
T_{c}$ \cite{KKS-PRL} becomes inaccurate. However, the results of the
present method are expected to be accurate at all $T$. Indeed, in the limit $%
T\gg T_{c}$ the present results (unlike \cite{KKS-PRL}) merge with those for
the ideal gas. This is physically appealing since at high $T$ the kinetic
energy becomes much larger than the interaction energy and the gas behaves
ideally. Similar to the ideal gas, the interacting mesoscopic BEC $\bar{n}%
_{0}(T)$ exhibits a smooth transition when passing through $T_{c}$.

One can see from Fig. \ref{n14box} that the repulsive interaction stimulates
BEC, and yields an increase in $\bar{n}_{0}$ at intermediate temperatures,
as compared to the ideal gas. This effect is known as \textquotedblleft
attraction in momentum space" and occurs for energetic reasons \cite{Legg01}%
. Bosons in different states interact more strongly than when they are in
the same state, and this favors multiple occupation of a single one-particle
state.

In conclusion, in this paper we presented a simple method which, for the
first time, yields an accurate description of the distribution function and
fluctuations for mesoscopic interacting Bogoliubov BEC in the canonical
ensemble at all temperatures. Our approach combines the analytical results
of \cite{KKS-PRL} with the laser-like master equation of \cite{Scul06}.

We gratefully acknowledge the support of the Office of Naval Research (Award
No. N00014-03-1-0385) and the Robert A. Welch Foundation (Grant No. A-1261).

\begin{widetext}
\onecolumngrid

%%%%%%%%%%%%%%%%%%%% Fig 3 %%%%%%%%%%%%%%%%%%%%%%%%%%%%%
\begin{figure}[h]
\bigskip
\centerline{\epsfxsize=1.1\textwidth\epsfysize=0.9\textwidth
\epsfbox{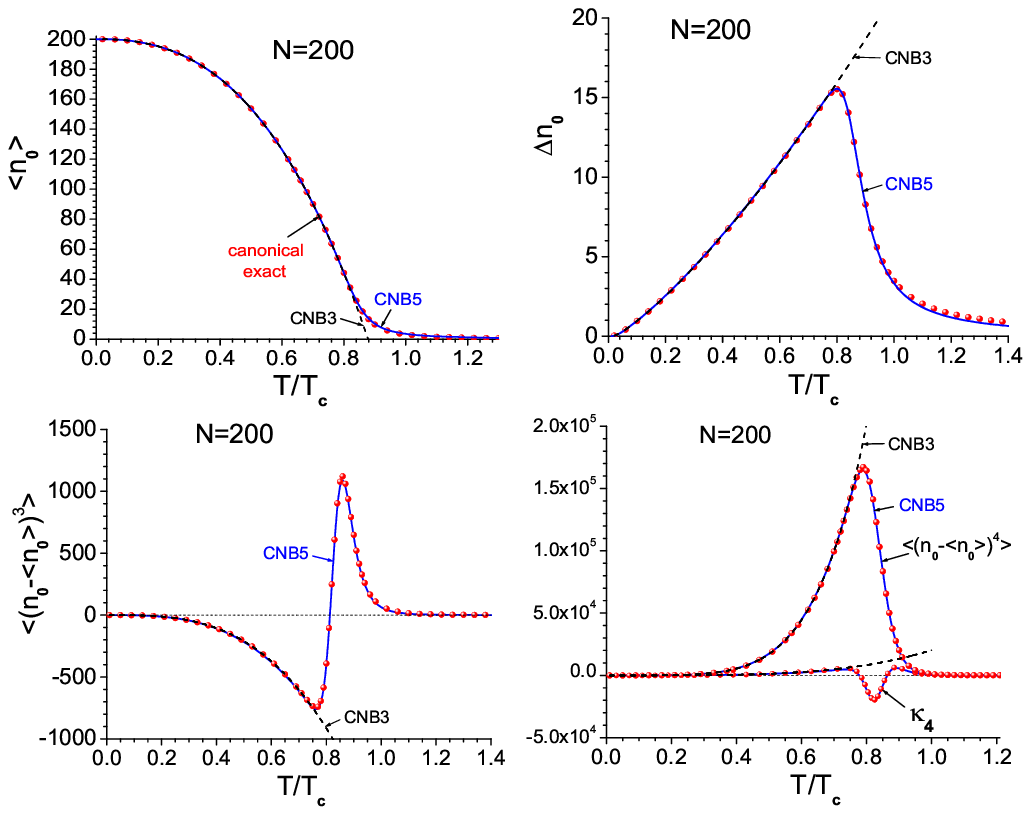}}
\caption{
Average condensate particle number $<n_0>$, its variance
$\Delta n_0=\sqrt{<(n_0-\bar n_0)^2>}$,
third and fourth central moments
$<(n_{0}-\bar{n}_{0})^{m}>$ ($m=3,4$)
and fourth cumulant
$\kappa_4$ as a
function of temperature for an ideal gas of $N=200$ particles in a harmonic
trap.
Solid lines (CNB5) show the result of the present approach.
\cite{KKS-PRL} yields
dashed lines (CNB3). Dots are ``exact" numerical simulation in the
canonical ensemble.
The temperature is normalized by the thermodynamic critical temperature
for the trap
$T_{c}=\hbar \omega N^{1/3}/k_{B}\zeta (3)^{1/3}$, where $\omega $ is the
trap frequency.
}
\label{n14trap}
\end{figure}
%%%%%%%%%%%%%%%%%%%%%%%%%%%%%%%%%%%%%%%%%%%%%%%%%%%%%%%%%

%%%%%%%%%%%%%%%%%%%% Fig 4 %%%%%%%%%%%%%%%%%%%%%%%%%%%%%
\begin{figure}[h]
\bigskip
\centerline{\epsfxsize=1.1\textwidth\epsfysize=0.9\textwidth
\epsfbox{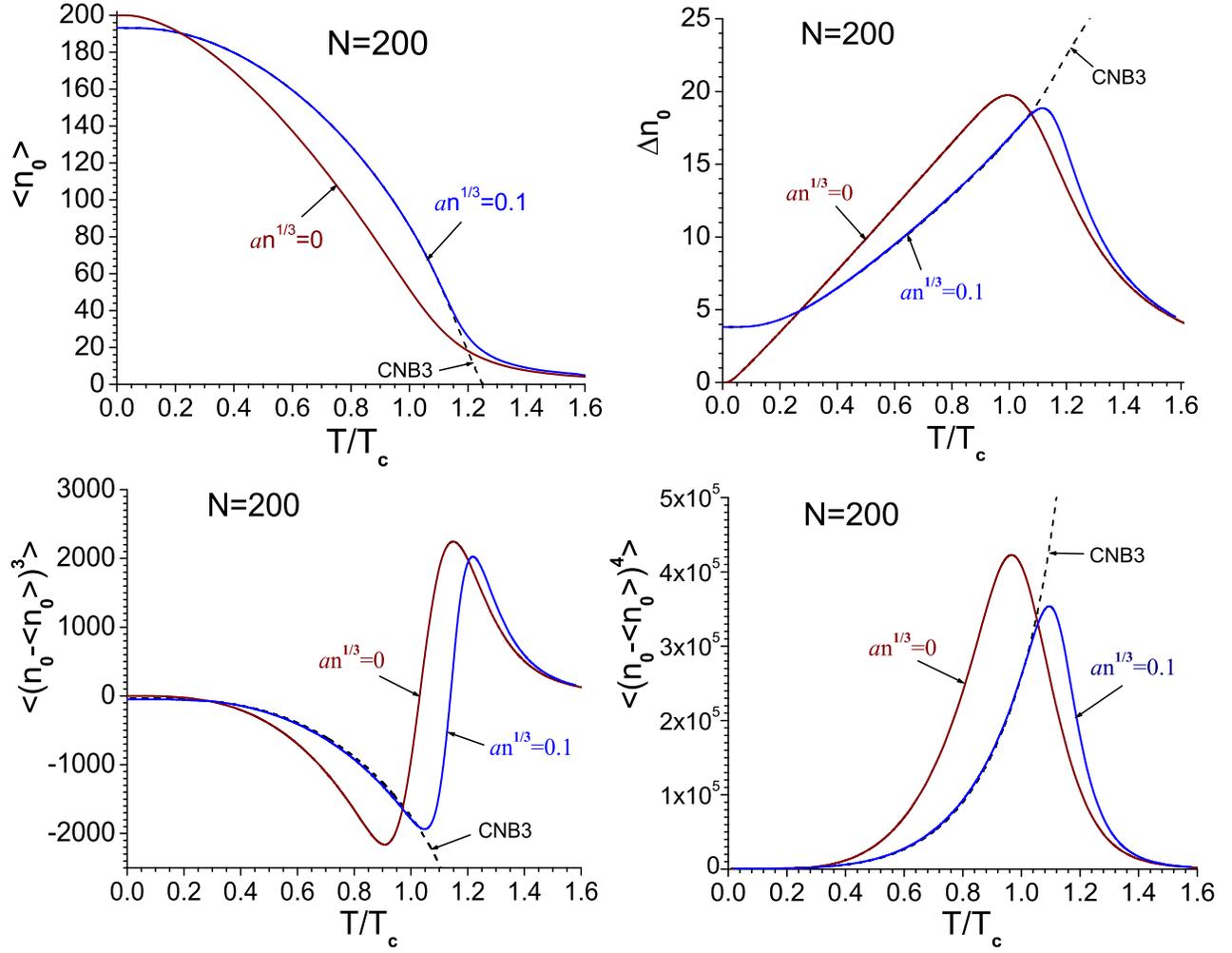}}
\caption{
Average condensate particle number, its variance,
third and fourth central moments
as a function of
temperature for an ideal ($an^{1/3}=0$) and interacting ($an^{1/3}=0.1$)
Bose gas
of $N=200$ particles in a box. Solid lines are the result of the present
approach. \cite{KKS-PRL} yields dashed lines (CNB3).
The temperature is normalized by the thermodynamic critical temperature
for the box
$T_{c}=2\pi
\hbar^2 n^{2/3}/k_{B}M\zeta (3/2)^{2/3}$, where $M$ is the
particle mass.
}
\label{n14box}
\end{figure}
%%%%%%%%%%%%%%%%%%%%%%%%%%%%%%%%%%%%%%%%%%%%%%%%%%%%%%%%%

\end{widetext}\twocolumngrid


\begin{thebibliography}{99}
\bibitem{kum} Cumulants $\kappa _{m}$ are defined as coefficients in Taylor
expansion $\ln \Theta _{n}(u)=\sum_{m=1}^{\infty }\kappa _{m}(iu)^{m}/m!$,
where $\Theta _{n}(u)$ is the characteristic function $\Theta
_{n}(u)=Tr\left\{ e^{iu\hat{n}}\hat{\rho}\right\} $. There are simple
relations between $\kappa _{m}$ and central moments $\mu _{m}$, in
particular, $\kappa _{1}=\bar{n}$, $\kappa _{2}=\mu _{2}$, $\kappa _{3}=\mu
_{3}$, $\kappa _{4}=\mu _{4}-3\mu _{2}^{2}$, $\kappa _{5}=\mu _{5}-10\mu
_{2}\mu _{3}$ and $\kappa _{6}=\mu _{6}-15\mu _{2}(\mu _{4}-2\mu _{2}^{2})$.
For Gaussian distribution $\kappa _{m}=0$, for $m=3,4,\ldots $.
\bibitem{note} The present paper is the fifth in the series of  our
Condensation of N Bosons papers.
\end{thebibliography}
\end{document}